\documentclass[12pt,a4paper,dvips]{article}
\usepackage{graphicx,epsfig}
\usepackage{hhline}

\usepackage{amsmath,amssymb}
\usepackage{times}
\usepackage[varg]{txfonts}
\DeclareMathAlphabet{\mathbold}{OML}{txr}{b}{it}

\usepackage{array,multirow,dcolumn}
\usepackage[mathlines,displaymath]{lineno}
\usepackage{rotating}
\usepackage[english]{babel}
 
\usepackage[numbers,square,comma,sort&compress]{natbib}
\usepackage{hypernat}
\usepackage{textcomp}
\newcolumntype{.}{D{.}{.}{-1}}
\newcolumntype{-}{D{-}{-}{-1}}

\usepackage[dvipsnames]{color}
\definecolor{rltred}{rgb}{0.75,0,0}
\definecolor{rltgreen}{rgb}{0,0.5,0}
\definecolor{rltblue}{rgb}{0,0,0.5}

\usepackage[hyperindex,bookmarks,bookmarksnumbered,breaklinks,a4paper,unicode]{hyperref}
\hypersetup{%
  pdftitle        = {EW2},
  urlcolor        = rltblue,       
  urlbordercolor  = 0 0 0.5,
  filecolor       = rltblue,       
  filebordercolor = 0 0 0.5,
  linkcolor       = rltred,        
  linkbordercolor = 0.75 0 0,
  citecolor       = rltgreen,      
  citebordercolor = 0 0.5 0,
  pagecolor       = rltgreen,      
  pagebordercolor = 0 0.5 0,
  menucolor       = rltgreen,      
  menubordercolor = 0 0.5 0,
  colorlinks    = true,
  pdfauthor     = {H1 Collaboration},
  pdfsubject    = { },
  pdfkeywords   = {High-Energy Physics, Particle Physics, Proton Structure, DIS}
}

\newlength{\dinwidth}
\newlength{\dinmargin}
\begin{document}

\makeatletter \def\NAT@space{} \makeatother

\begin{titlepage}
 
\noindent
DESY-16-063

\vspace*{2.5cm}

\begin{center}
\begin{Large}

{\bfseries Combined Electroweak and QCD Fit to HERA Data}

\vspace*{2cm}

I.~Abt$^a$,
A.M.~Cooper-Sarkar$^b$,
B.~Foster$^{b,c,d}$, 
C.~Gwenlan$^b$,
V.~Myronenko$^d$,
O.~Turkot$^d$,
K.~Wichmann$^d$, 

\end{Large}
\end{center}

{\protect \hskip 0.5cm} $^a$ Max-Planck-Institut f{\"u}r Physik, Werner-Heisenberg-Institut, M{\"u}nchen, Germany 

\vspace*{-.12cm}
{\protect \hskip 0.5cm} $^b$ Physics Department, University of Oxford, Oxford, U.K.
 
\vspace*{-.12cm}
{\protect \hskip 0.5cm} $^c$ Hamburg University, I. Institute of Exp. Physics, Hamburg, Germany

\vspace*{-.12cm}
{\protect \hskip 0.5cm} $^d$ Deutsches Elektronen Synchrotron DESY, Hamburg, Germany


\vspace*{1cm}

\begin{abstract} \noindent

A simultaneous Electroweak and QCD fit of 
electroweak parameters
and parton distribution functions  
to HERA data on deep inelastic scattering is presented. 
The input data are neutral current and charged current  
inclusive cross sections measured by the H1 and ZEUS
collaborations at the $ep$ collider HERA. 
The polarisation of the electron beam was taken into account 
for the ZEUS and H1 data recorded between 2004 and 2007. Results 
are presented on the vector and axial-vector couplings of the $Z$ boson to 
$u$- and $d$-type quarks. 
The values are in agreement with 
Standard Model predictions.
The results on $a_u$ and $v_u$ represent the most precise measurements from
a single process.

\end{abstract}

\vspace*{1.5cm}

\end{titlepage}

\newpage
~~
\newpage

\section{Introduction}
\label{sec:intro}

Data on deep inelastic electron\footnote{In this paper, 
the word ``electron'' refers to both electrons and positrons, 
unless otherwise stated.}--proton, $ep$, scattering (DIS) 
have been used in analyses within the framework of Quantum Chromo Dynamics (QCD)
for many years~\cite{MANDY} and have formed the basis of investigations
of the structure of the proton.
The data from the $ep$ collider HERA extended the reach in the 
four-momentum-transfer squared, $Q^2$, and in Bjorken $x$, $x_{Bj}$, by 
several orders of magnitude with respect to previous fixed-target 
experiments~\cite{rmp:71:1275}. 
At HERA, values of $Q^2$ of up to $50\,000\,$GeV$^2$ were reached,
a regime where 
the contribution of $Z$ exchange becomes comparable
to the contribution from 
photon exchange.

During the HERA\,II running period, the HERA collider provided a significant
amount of data with beams longitudinally polarised 
to an average level between 25\,\% and 35\,\%. 
This facilitates detailed studies of electroweak (EW) effects.
Recently, the ZEUS collaboration published a combined
QCD and electroweak analysis~\cite{ZEUS-EW} exploiting the
ZEUS neutral current (NC) and charged current (CC) 
$e^+p$ and $e^-p$ inclusive cross sections 
for polarised 
beams~\cite{ZEUS2NCe,ZEUS2NCp,ZEUS2CCe,ZEUS2CCp}. 
For the analysis presented here, cross sections published
by the H1 collaboration~\cite{H1allhQ2} for polarised beams were
also considered.
These data sets, together with data sets for unpolarised 
beams originally published by
H1~\cite{H1FL1,H1FL2,Adloff:1999ah,Adloff:2000qj,Adloff:2003uh,Collaboration:2009kv,Collaboration:2009bp}
and ZEUS~\cite{ZEUSFL,Breitweg:1998dz,zeuscc97,Chekanov:2001qu,Chekanov:2002zs,Chekanov:2002ej,Chekanov:2003vw,Chekanov:2003yv}~\footnote{As used as input to the data combination presented by the
H1 and ZEUS collaborations~\cite{HERAPDF20}.},
were used as input to a combined QCD and EW fit, HH-EW-Z.
This fit was used to determine
the couplings of the $Z$ boson to $u$- and $d$-type quarks.

\section{QCD and EW Combined Analysis}

The analysis presented here follows closely the method described
in detail in the ZEUS publication~\cite{ZEUS-EW}. 
It uses the next-to-leading-order (NLO)
DGLAP~\cite{Gribov:1972ri,Gribov:1972rt,Lipatov:1974qm,Dokshitzer:1977sg,Altarelli:1977zs}
formalism to describe the evolution of the 
parton distribution functions (PDFs) with $Q^2$
and
the on-shell definition of 
$\sin^2\theta_W = 1 - M_W^2/M_Z^2$,
where $\sin^2\theta_W$ is the electroweak mixing angle,
and $M_W$ and $M_Z$ are the mass of the $W$ and $Z$ boson, respectively.
The EW part of the analysis was performed at leading order with
partial higher-order corrections in the on-shell scheme.
The RT variable-number heavy-flavour 
scheme~\cite{Thorne:1997ga,Thorne:2006qt,Thorne:RTopt} was employed and
the values of PDG14~\cite{PDG14} were used 
for all masses and couplings throughout the analysis, unless they were free 
parameters in a fit.

The PDFs of the proton were
parameterised with 13 free parameters as
\begin{eqnarray}
\label{eq:xgpar}
xg(x) &=   & A_g x^{B_g} (1-x)^{C_g} - A_g' x^{B_g'} (1-x)^{C_g'}  ,  \\
\label{eq:xuvpar}
xu_v(x) &=  & A_{u_v} x^{B_{u_v}}  (1-x)^{C_{u_v}}\left(1+E_{u_v}x^2 \right) , \\
\label{eq:xdvpar}
xd_v(x) &=  & A_{d_v} x^{B_{d_v}}  (1-x)^{C_{d_v}} , \\
\label{eq:xubarpar}
x\bar{U}(x) &=  & A_{\bar{U}} x^{B} (1-x)^{C_{\bar{U}}} , \\
\label{eq:xdbarpar}
x\bar{D}(x) &= & A_{\bar{D}} x^{B} (1-x)^{C_{\bar{D}}} ,
\end{eqnarray}
where $x$ is the fraction of the proton momentum carried by the quark.
The normalisation parameters, 
$A_{u_v}, A_{d_v}, A_g$, are constrained 
by the quark-number sum rules and the momentum sum rule. 
The strange-quark distribution is expressed 
as an $x$-independent fraction, $f_s$, of the $d$-type sea, 
$x\bar{s}= 0.4\, x\bar{D}$ at the 
starting scale $\mu^2_{{\rm f}_0}=1.9$\,GeV$^2$.
The parameter ${C_g'}$ is fixed to  ${C_g'} = 25$~\cite{Martin:2009iq}.

The PDF parameters were fitted to the HERA 
inclusive cross sections together with the axial-vector and vector
couplings of the
$Z$ boson to the $u$- and $d$-type quarks, $a_u$, $a_d$, $v_u$ and $v_d$,
respectively.
For this fit, called HH-EW-Z,
the ZEUSfitter package\,\footnote{The package was also used
in  the combined ZEUS electroweak and QCD analysis~\cite{ZEUS-EW}.} 
was used. 
The results were cross-checked with the HERAFitter~\cite{HERAFitter} package.

All cross sections for unpolarised beams were used as originally published
by H1~\cite{H1FL1,H1FL2,Adloff:1999ah,Adloff:2000qj,Adloff:2003uh,Collaboration:2009kv,Collaboration:2009bp}
and ZEUS~\cite{ZEUSFL,Breitweg:1998dz,zeuscc97,Chekanov:2001qu,Chekanov:2002zs,Chekanov:2002ej,Chekanov:2003vw,Chekanov:2003yv}.
The H1 cross sections for polarised beams
were also used as published by H1~\cite{H1allhQ2}.
The ZEUS cross sections for
polarised beams~\cite{ZEUS2NCe,ZEUS2NCp,ZEUS2CCe,ZEUS2CCp} were used
as originally published, but with updated values of the polarisation
as published in the ZEUS EW analysis~\cite{ZEUS-EW}.
In addition, for the present analysis, extra uncertainties 
were added to the uncorrelated
systematic uncertainties on these ZEUS data.
In their original publications the ZEUS collaboration did not consider 
systematic uncertainties
on EW corrections, whereas the H1 collaboration included such 
uncertainties~\cite{H1allhQ2}. The uncertainties now added to the ZEUS data are
equivalent to the uncertainties on the EW corrections included by H1.

\section{The HH-EW-Z Fit and the $ \mathbf{Z} $  Couplings}

\begin{figure}[p]
\vfill
\begin{center}
\includegraphics[width=6in]{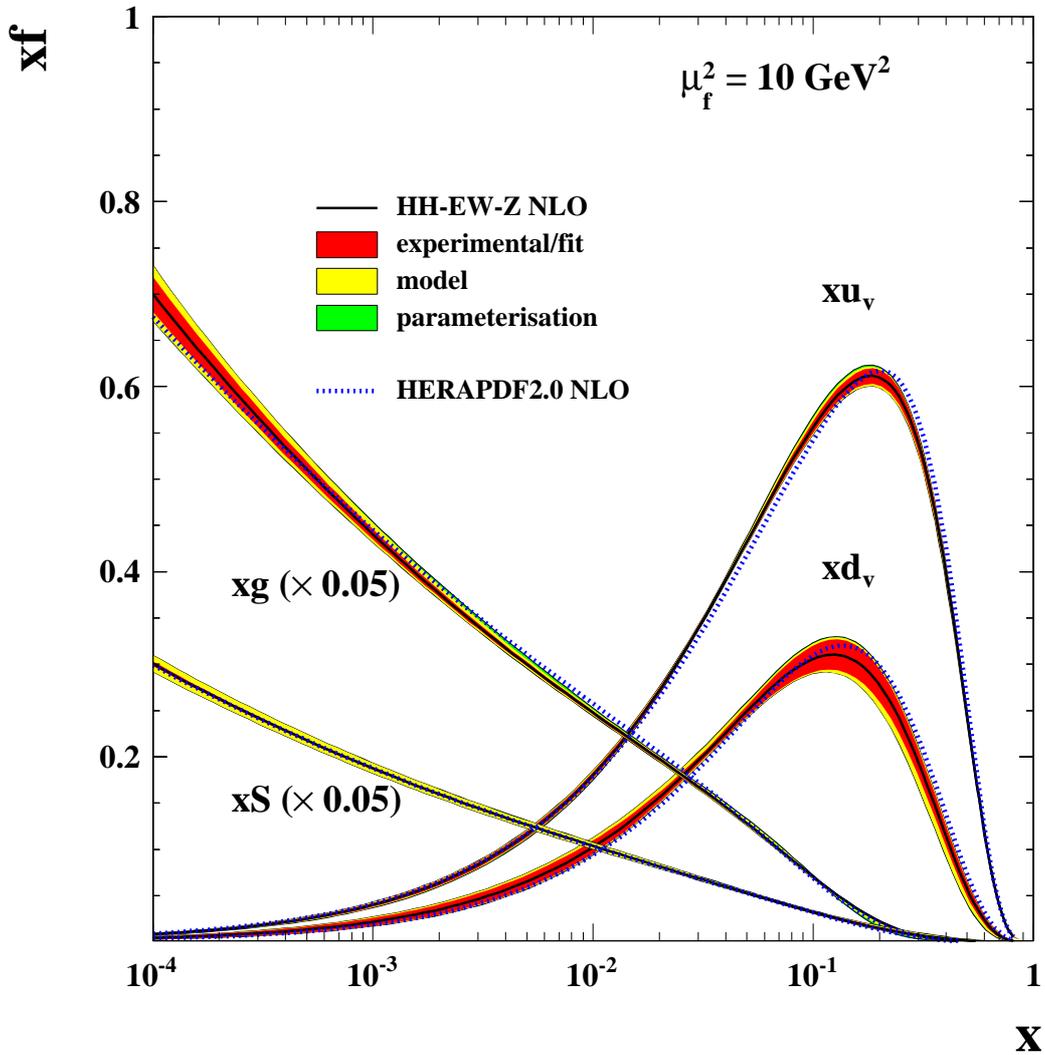}
\end{center}
\caption{The NLO PDF set HH-EW-Z with
         cumulative  
         experimental/fit, model and
         parameterisation uncertainties at the 
         factorisation scale $\mu_{\rm f}^2=10$\,GeV$\,^2$.
         All positive and negative uncertainties 
         in the model were added separately in quadrature.
         The parameterisation uncertainty represents
         an envelope of all individual parameterisation
         uncertainties.  
         Also shown are the 
         central values of HERAPDF2.0 NLO.         
}
\label{fig:HH-EW-Z}
\vfill
\end{figure}

The PDFs of the  HH-EW-Z fit are shown in Fig.~\ref{fig:HH-EW-Z}
with experimental/fit, model and parameterisation
uncertainties, determined according to the prescriptions
of the HERAPDF2.0 analysis~\cite{HERAPDF20}.
Also shown are the central values 
for the PDFs of HERAPDF2.0 NLO.
The PDFs are very similar.
The PDF parameters of HH-EW-Z are only weakly correlated
to the $Z$ couplings. The full correlation matrix for the
13 PDF parameters and the four $Z$ couplings is given in
Table~\ref{tab:matrix}.

\begin{figure}[p]
\vfill
\begin{center}
\includegraphics[width=6in]{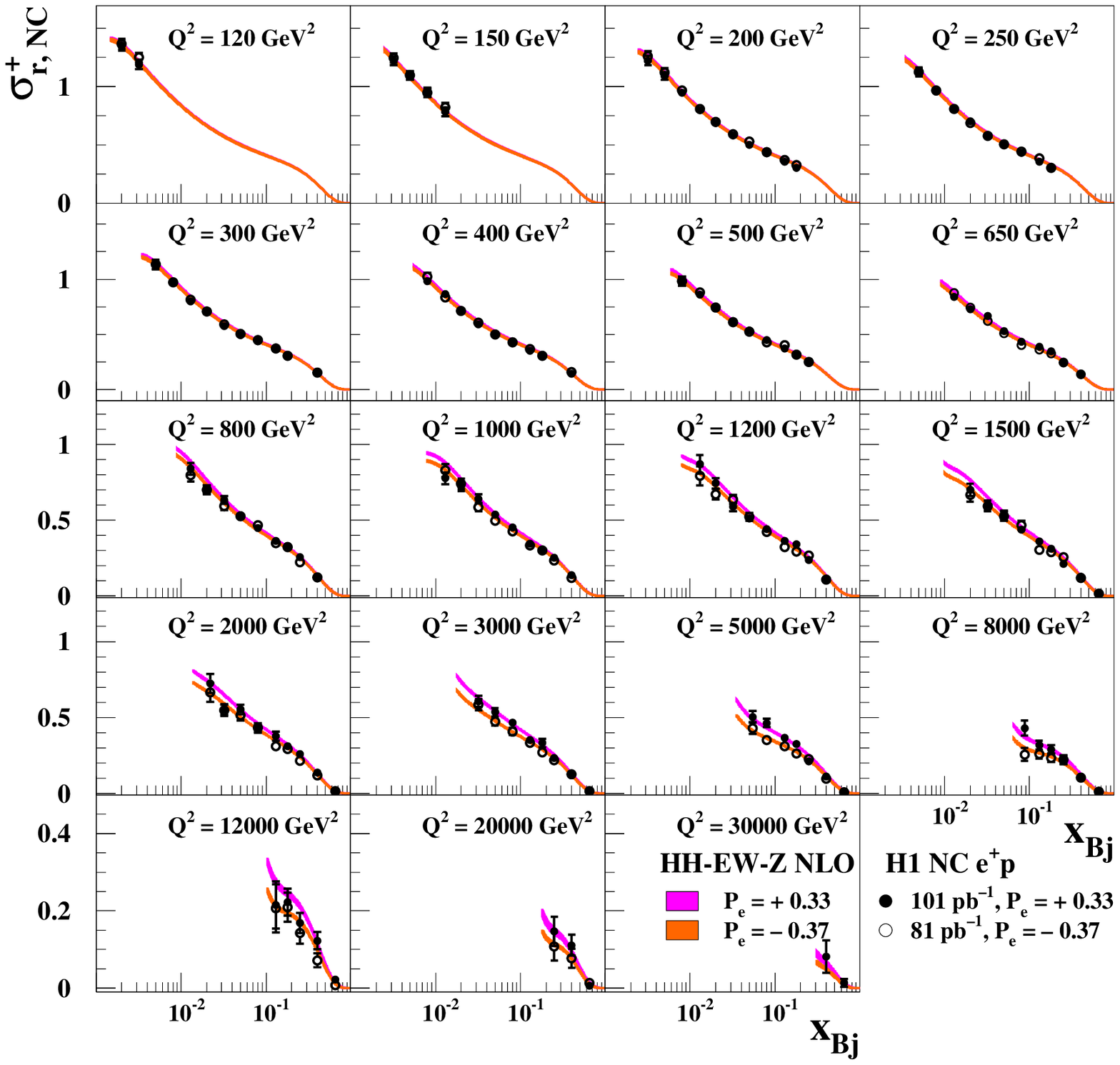}
\end{center}
\caption{The NLO predictions of HH-EW-Z compared to 
  the H1 $e^+p$ NC DIS reduced cross-sections $\sigma^+_{r,NC}$
  for positively and negatively polarised beams
  plotted as a function of $x_{\rm Bj}$ at fixed values of $Q^2$.
  The closed (open) circles represent the H1 data
  for positive (negative) polarisation.
  The bands indicate the full uncertainties on the 
  predictions of HH-EW-Z.
}
\label{fig:NC:H1}
\vfill
\end{figure}

\begin{figure}[p]
\vfill
\begin{center}
\includegraphics[width=6in]{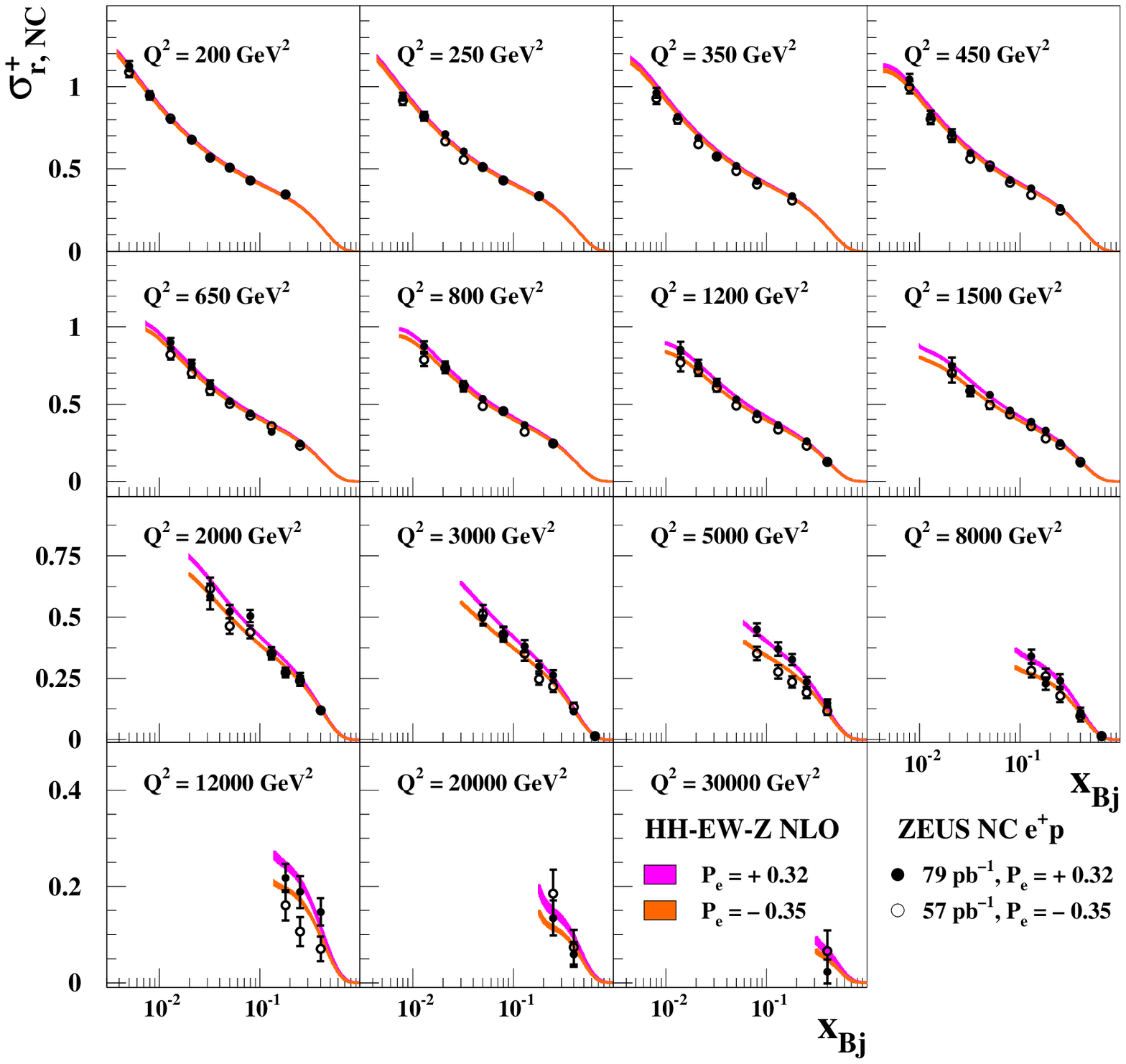}
\end{center}
\caption{The NLO predictions of HH-EW-Z compared to 
  the ZEUS $e^+p$ NC DIS reduced cross-sections $\sigma^+_{r,NC}$
  for positively and negatively polarised beams
  plotted as a function of $x_{\rm Bj}$ at fixed values of $Q^2$.
  The closed (open) circles represent the ZEUS data
  for positive (negative) polarisation.
  The bands indicate the full uncertainties on the 
  predictions of HH-EW-Z.
}
\label{fig:NC:Z}
\vfill
\end{figure}

The $\chi^2$ per degree of freedom for HH-EW-Z is 3556/3231=1.10.
This can be compared to 1.12 for ZEUS-EW-Z~\cite{ZEUS-EW}
and 1.20 for HERAPDF2.0 NLO~\cite{HERAPDF20}.
The description of the data is very good.
The predictions of HH-EW-Z are compared to the
high-precision $e^+p$ NC data from H1~\cite{H1allhQ2} 
and ZEUS~\cite{ZEUS2NCp}
in Figs.~\ref{fig:NC:H1} and~\ref{fig:NC:Z}, respectively.

The result of HH-EW-Z for the couplings of the $Z$ boson to $u$- and
$d$-type quarks are
\begin{eqnarray}
\nonumber
a_u & = & +0.532~ ^{+0.081}_{-0.058} \,{\rm {\scriptstyle{(experimental/fit)}}}~ 
        ^{+0.036}_{-0.022}\,{\rm {\scriptstyle{(model)}}}~ 
        ^{+0.060}_{-0.008}\,{\rm {\scriptstyle{(parameterisation)}}} ~,\\
\nonumber
a_d & = & -0.409~ ^{+0.327}_{-0.199}\,{\rm {\scriptstyle{(experimental/fit)}}}~ 
        ^{+0.112}_{-0.071}\,{\rm {\scriptstyle{(model)}}}~ 
        ^{+0.140}_{-0.026}\,{\rm {\scriptstyle{(parameterisation)}}} ~,\\
\nonumber
v_u & = & +0.144~ ^{+0.065}_{-0.050}\,{\rm {\scriptstyle{(experimental/fit)}}}~ 
        ^{+0.013}_{-0.014}\,{\rm {\scriptstyle{(model)}}}~
        ^{+0.002}_{-0.025}\,{\rm {\scriptstyle{(parameterisation)}}}  ~,\\
\nonumber
v_d & = & -0.503~ ^{+0.168}_{-0.093}\,{\rm {\scriptstyle{(experimental/fit)}}}~ 
        ^{+0.031}_{-0.028}\,{\rm {\scriptstyle{(model)}}}~ 
        ^{+0.006}_{-0.036}\,{\rm {\scriptstyle{(parameterisation)}}} ~.
\end{eqnarray}
These values are compared to the results from ZEUS-EW-Z~\cite{ZEUS-EW} 
in Table~\ref{tab:Z}.
They agree within uncertainties. 
Also listed are  SM predictions
and 
values obtained from fits which were performed as cross-checks:
\begin{itemize}
\item a fit with the PDFs fixed to those of a 13-parameter QCD-only
      fit, HH-13p;
\item a fit with the PDFs fixed to those of HERAPDF2.0.
\end{itemize}
Only experimental/fit uncertainties
were considered for these cross checks.
The values agree within the experimental uncertainties
with the result from HH-EW-Z.

\begin{figure}[p]
\vfill
\begin{center}
\includegraphics[width=4in]{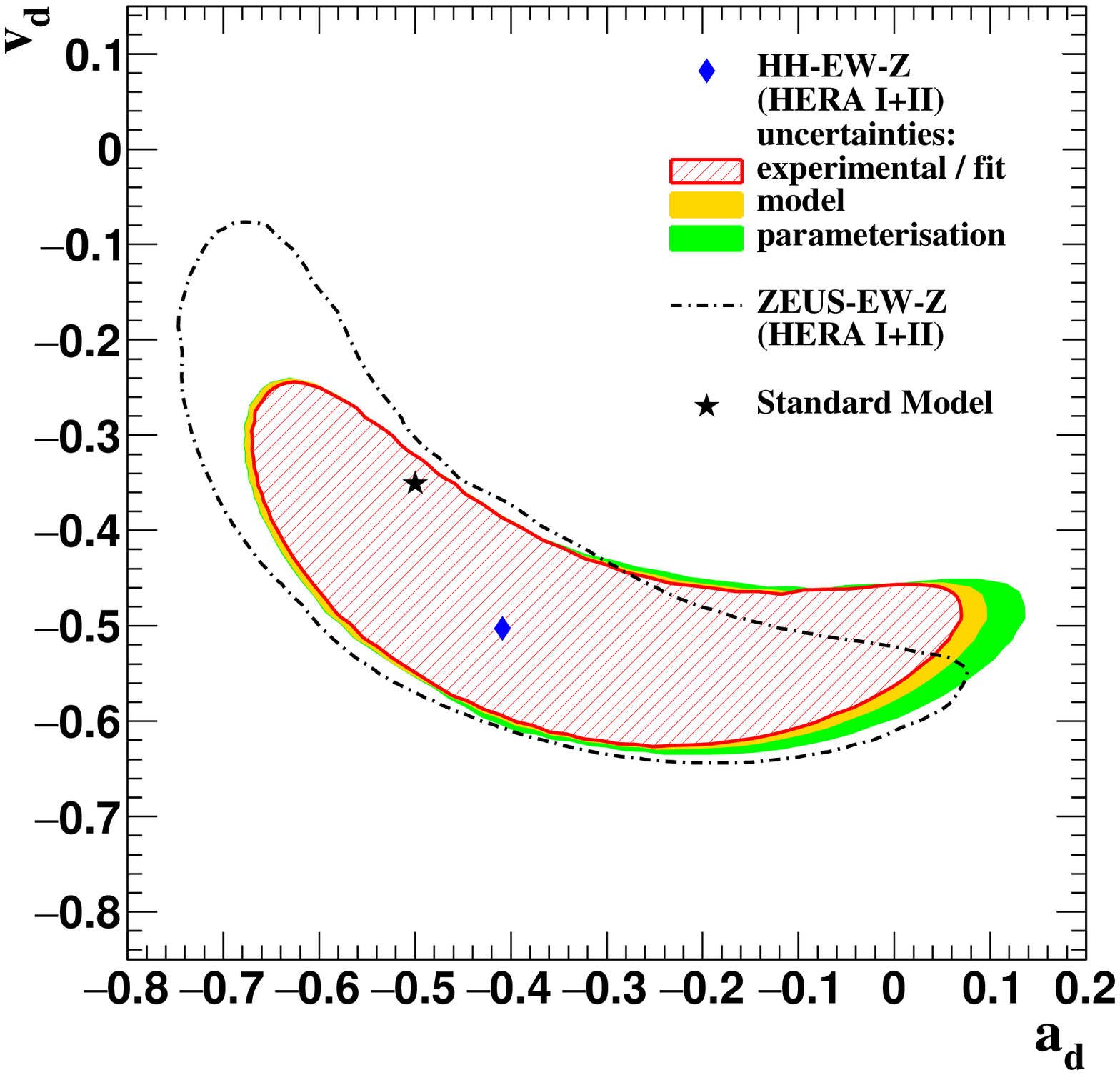}
\includegraphics[width=4in]{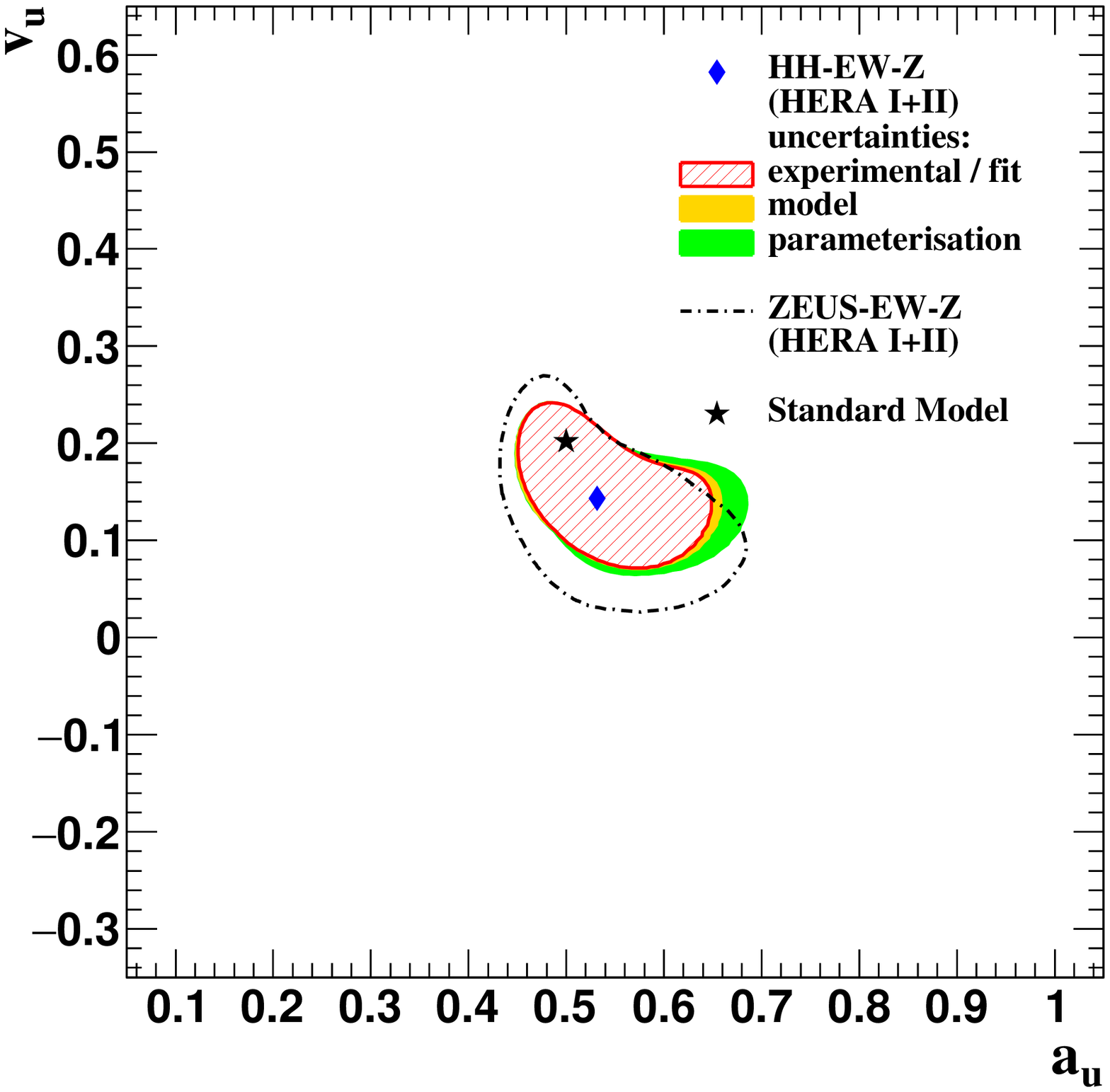}
\end{center}
\caption{The 68\,\%\,C.L. contours 
         for $a_d,v_d$ and $a_u,v_u$
         obtained for the HH-EW-Z fit.
         Also shown are the 68\,\%\,C.L. contours for
         the ZEUS-EW-Z fit with total uncertainties.
}
\label{fig:av-us}
\vfill
\end{figure}

\begin{figure}[p]
\vfill
\begin{center}
\includegraphics[width=4in]{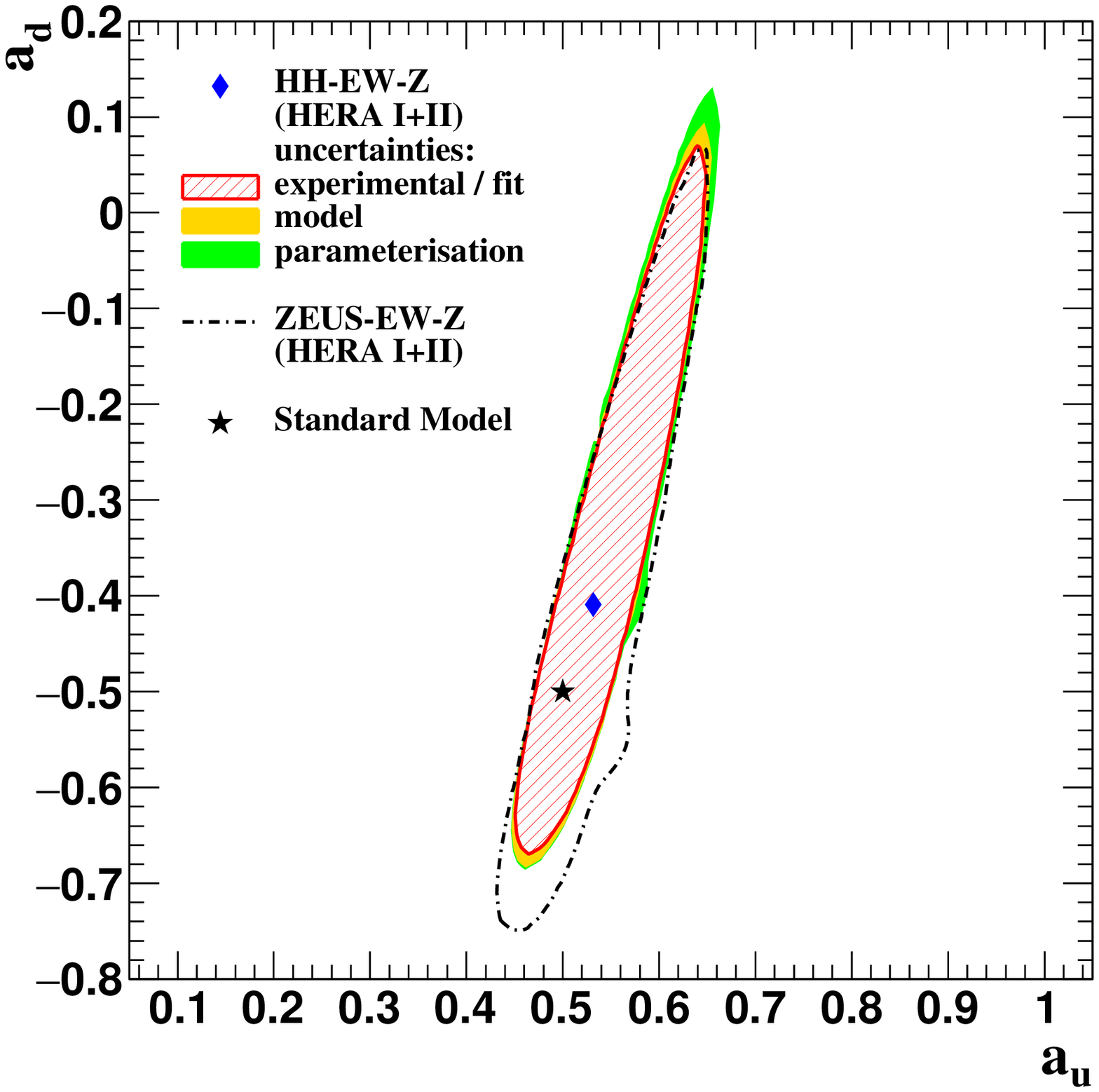}
\includegraphics[width=4in]{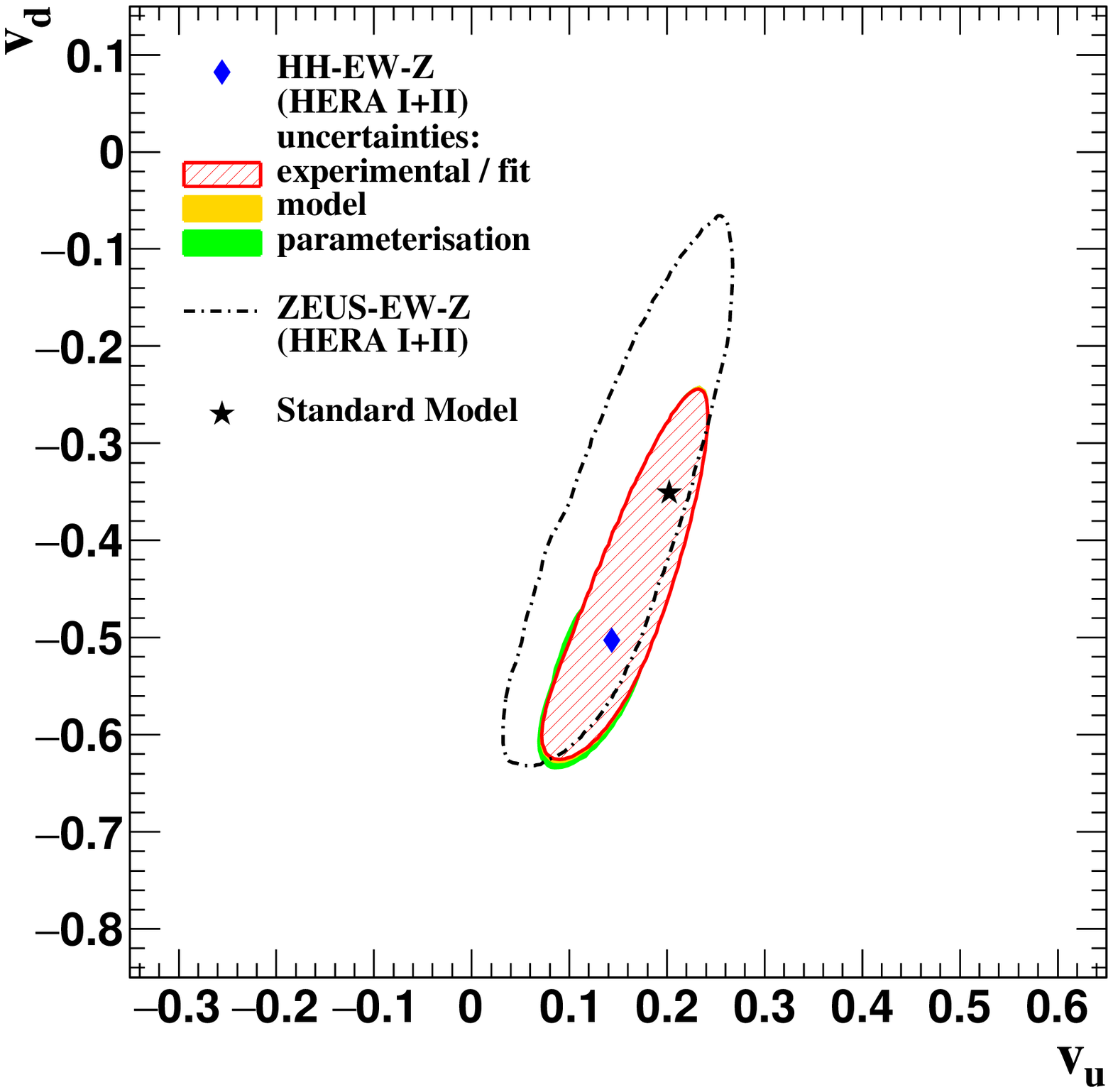}
\end{center}
\caption{The 68\,\%\,C.L. contours 
         for $a_u,a_d$ and $v_u,v_d$
         obtained for the HH-EW-Z fit.
         Also shown are the 68\,\%\,C.L. contours for
         the ZEUS-EW-Z fit with total uncertainties.
}
\label{fig:aavv}
\vfill
\end{figure}

\begin{figure}
\vfill
\begin{center}
\includegraphics[width=4in]{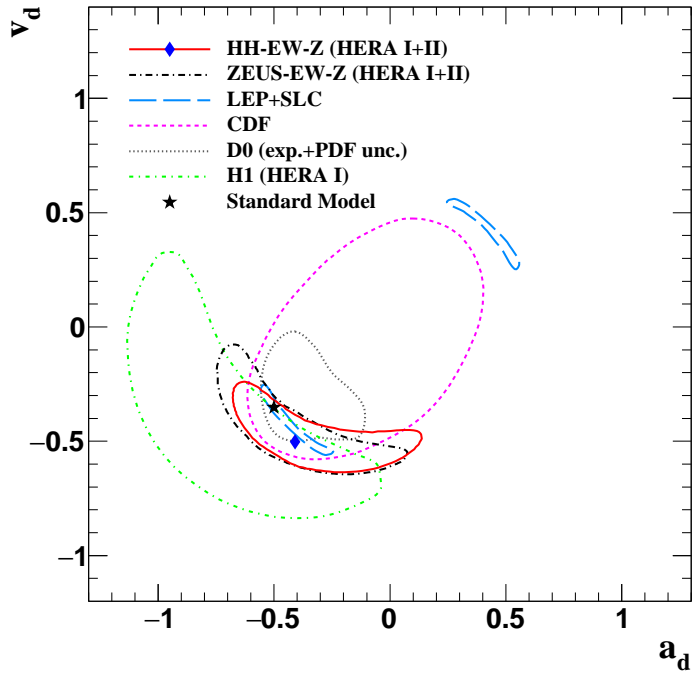}
\includegraphics[width=4in]{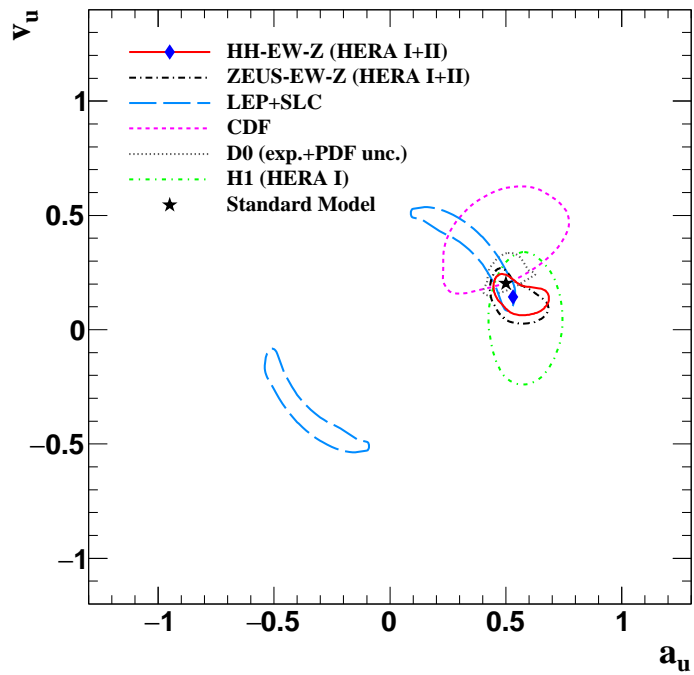}
\end{center}
\caption{The 68\,\%\,C.L. contours 
         for $a_d,v_d$ and $a_u,v_u$
         obtained for the HH-EW-Z fit.
         Also shown are  
         results 
         from ZEUS-EW-Z, HERA\,I (H1), 
         LEP (ALEPH, OPAL, L3 and DELPHI) 
         plus SLC (SLD) combined, 
         and the Tevatron (CDF and D0). 
}
\label{fig:av-all}
\vfill
\end{figure}

\begin{figure}[p]
\vfill
\begin{center}
\includegraphics[width=6in]{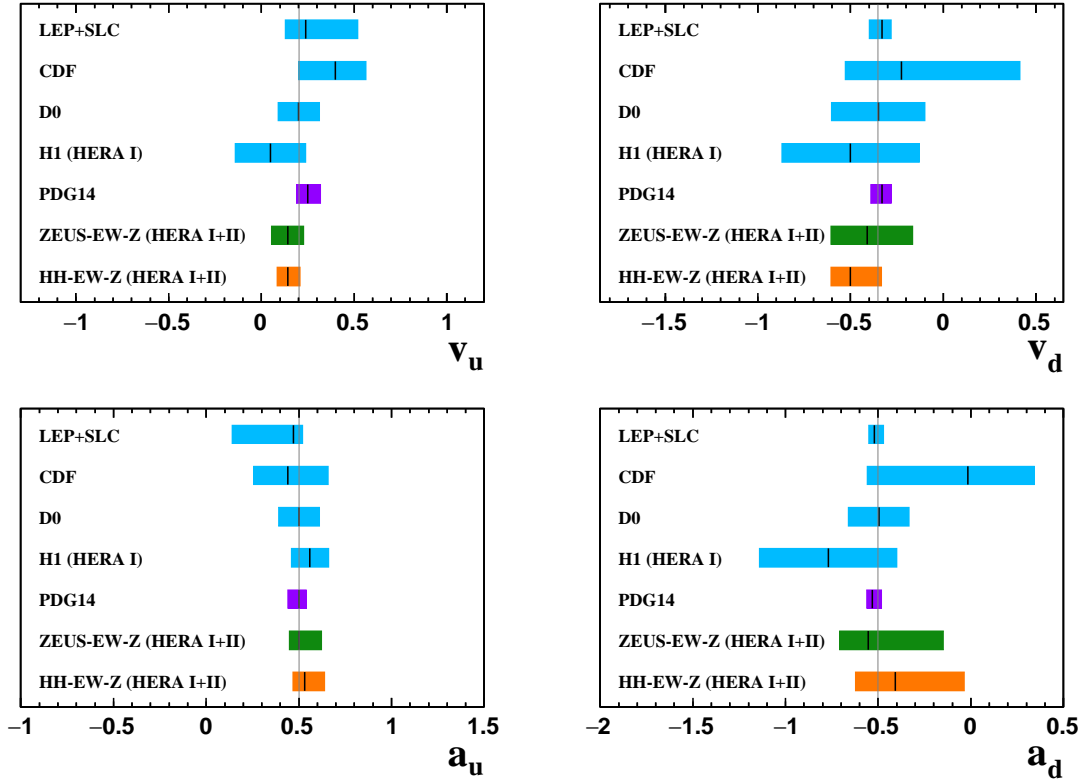}
\end{center}
\caption{The values from the HH-EW-Z fit
         for $a_d, a_u, v_d$ and $v_u$ compared to
         the values from ZEUS-EW-Z and the 
         results from LEP (ALEPH, OPAL, L3 and DELPHI) 
         plus SLC (SLD) combined, 
         the Tevatron (CDF and D0), 
         HERA\,I (H1). 
         The PDG14 world average is also shown; this does
         not contain the measurements 
         from the HH-EW-Z and ZEUS-EW-Z fits
         based on all HERA data. 
         All results are given with
         total uncertainties.
         Vertical black lines in each box 
         indicate central values,
         the long gray vertical lines indicate the SM
         predictions. 
}
\label{fig:comparison}
\vfill
\end{figure}

Profile likelihood contours at 68\,\%\,C.L. 
for the couplings
were obtained as described in the
ZEUS publication~\cite{ZEUS-EW}.
They are
shown~\footnote{Numerical information 
is available as additional material for this publication.} 
for $a_u,v_u$ and $a_d,v_d$ 
in Fig.~\ref{fig:av-us}                     
and for $a_u,a_d$ and $v_u,v_d$ in
Fig.~\ref{fig:aavv}.
These figures demonstrate
very clearly that the HERA data constrain the couplings of the $Z$ boson to 
the $u$ quark significantly better
than the couplings to the $d$ quark.
This is due to the larger $u$ valence content of the proton and
the larger charge of the $u$ quark. 
The couplings as determined by HH-EW-Z are compatible with
the SM.
Figure~\ref{fig:av-all} shows
the 68\,\%\,C.L. contours from HH-EW-Z, together
with the contours from ZEUS-EW-Z~\cite{ZEUS-EW}
and the measurements from LEP+SLC~\cite{ZLEPSLC}, 
the Tevatron~\cite{ZD0Old,ZCDF} and HERA\,I (H1)~\cite{ZH1}.
The fits HH-EW-Z and ZEUS-EW-Z are based
both on HERA\,I and HERA\,II data and were not included in the
combinations for PDG14~\cite{PDG14}. 
The PDG values and all measurements
are compared in Fig.~\ref{fig:comparison}.
The HH-EW-Z results on the axial-vector 
and vector couplings to $u$-type quarks 
are the most precise results published from a single process.
The vector couplings from HH-EW-Z are
significantly more accurate than from ZEUS-EW-Z.
This reflects the importance of the information
on the polarisation of the beams for the vector
couplings~\cite{ZEUS-EW}. 
Thus, the inclusion of the H1 data
for polarised beams is the reason for 
the improvement in these couplings.

The ZEUS collaboration also presented~\cite{ZEUS-EW} measurements of
the electroweak mixing angle and $M_W$.
These results do not depend strongly on the beam polarisation.
Two fits were performed as cross-checks with the 
13 PDF parameters fixed and either $\sin^2\theta_W$ 
or $M_W$ as free parameters.
The results are compatible with those of the ZEUS EW fits within 
experimental/fit uncertainties:
\begin{eqnarray}
\nonumber
\sin^2\theta_W & = 
      0.2255~ \pm 0.0011 ~{\rm {\scriptstyle{(experimental/fit)}}} & 
      {\rm HH\,EW}~, \\
\nonumber
\sin^2\theta_W & =  
      0.2252~ \pm 0.0011 ~{\rm {\scriptstyle{(experimental/fit)}}}  &
      {\rm ZEUS\,EW}~, \\
\nonumber
\\
\nonumber
  M_W & = 
      (80.74~ \pm 0.28 ~{\rm {\scriptstyle{(experimental/fit)}}})~{\rm GeV} 
           & {\rm HH\,EW}~, \\
\nonumber
  M_W & =  
      (80.68~ \pm 0.28 ~{\rm {\scriptstyle{(experimental/fit)}}})~{\rm GeV} 
          &{\rm ZEUS\,EW}~. 
\end{eqnarray}

A simultaneous fit to the 13 PDF parameters and both
$\sin^2\theta_W$ and $M_W$ 
also yielded results compatible with the results presented 
by ZEUS~\cite{ZEUS-EW}.
Since the sensitivity with respect to the ZEUS EW fits 
was not significantly increased, 
the detailed studies on $\sin^2\theta_W$ and $M_W$ 
presented in the ZEUS paper were not repeated.

%
\section{Summary and Conclusions \label{sec:sum}}

The results of a combined electroweak and QCD
fit to all available HERA inclusive DIS cross sections, 
taking into account beam polarisation for both the H1 and ZEUS data, 
have been presented.
The results on the couplings of the $Z$ boson to
$u$- and $d$-type quarks are:
\begin{eqnarray}
\nonumber
a_u & = & +0.532~ ^{+0.081}_{-0.058} \,{\rm {\scriptstyle{(experimental/fit)}}}~ 
        ^{+0.036}_{-0.022}\,{\rm {\scriptstyle{(model)}}}~ 
        ^{+0.060}_{-0.008}\,{\rm {\scriptstyle{(parameterisation)}}} ~,\\
\nonumber
a_d & = & -0.409~ ^{+0.327}_{-0.199}\,{\rm {\scriptstyle{(experimental/fit)}}}~ 
        ^{+0.112}_{-0.071}\,{\rm {\scriptstyle{(model)}}}~ 
        ^{+0.140}_{-0.026}\,{\rm {\scriptstyle{(parameterisation)}}} ~,\\
\nonumber
v_u & = & +0.144~ ^{+0.065}_{-0.050}\,{\rm {\scriptstyle{(experimental/fit)}}}~ 
        ^{+0.013}_{-0.014}\,{\rm {\scriptstyle{(model)}}}~
        ^{+0.002}_{-0.025}\,{\rm {\scriptstyle{(parameterisation)}}}  ~,\\
\nonumber
v_d & = & -0.503~ ^{+0.168}_{-0.093}\,{\rm {\scriptstyle{(experimental/fit)}}}~ 
        ^{+0.031}_{-0.028}\,{\rm {\scriptstyle{(model)}}}~ 
        ^{+0.006}_{-0.036}\,{\rm {\scriptstyle{(parameterisation)}}} ~.
\end{eqnarray}

These results are compatible with the Standard Model.
The exploitation of all available data for polarised beams
provides very accurate determinations of the $Z$-boson couplings.
The couplings to the $u$-type quarks are the most precise 
values published for a single process.

\section{Acknowledgments}

We are grateful to our ZEUS and H1 colleagues 
who supported this work. We especially thank Achim Geiser
for the many fruitful discussions.
We thank our funding agencies, especially the
Humboldt foundation and the Max-Planck-Society, 
for financial support and DESY for the hospitality extended 
to the non-DESY authors.

\clearpage
\bibliography{new}
  
\clearpage
%

\begin{sidewaystable}
\begin{center}
\begin{scriptsize}\renewcommand\arraystretch{1.1}
\begin{tabular}[H]{ | c | r r r r r r r r r r r r r r r r r |}
\hline
\hline
Parameters & {\it xg: B}       & {\it xg: C}   & {\it xg: A$^{'}$}    & {\it xg: B$^{'}$}    & {\it xu$_{v}$: B}   & {\it xu$_{v}$: C}  & {\it xu$_{v}$: E}    & {\it xd$_{v}$: B}    & {\it xd$_{v}$: C}  & {\it x$\bar{U}$: C} & {\it x$\bar{D}$: A}  & {\it x$\bar{D}$: B} & {\it x$\bar{D}$: C} & $a_{u}$ & $a_{d}$ & $v_{u}$ &  $v_{d}$ \rule{0pt}{3ex}  \\
\hline \rule{0pt}{3ex}

{\it xg: B}          & 1.000 & 0.491 & $-$0.224 & 0.935 & 0.012 & 0.106 & 0.044 & $-$0.049 & $-$0.078 & $-$0.049 & $-$0.098 & $-$0.140 & 0.018 & 0.057 & 0.061 & $-$0.039 & $-$0.051 \\
{\it xg: C}          & 0.491 & 1.000 & 0.660 & 0.707 & 0.287 & $-$0.267 & $-$0.464 & $-$0.054 & 0.196 & $-$0.047 & $-$0.140 & $-$0.175 & $-$0.369 & 0.106 & 0.093 & $-$0.124 & $-$0.114 \\
{\it xg: A$^{'}$}    & $-$0.224 & 0.660 & 1.000 & 0.125 & 0.513 & $-$0.361 & $-$0.593 & 0.226 & 0.254 & 0.162 & 0.084 & 0.072 & $-$0.100 & $-$0.038 & 0.003 & $-$0.065 & $-$0.070 \\
{\it xg: B$^{'}$}    & 0.935 & 0.707 & 0.125 & 1.000 & 0.200 & $-$0.002 & $-$0.144 & 0.048 & $-$0.008 & 0.042 & $-$0.017 & $-$0.056 & 0.018 & 0.033 & 0.057 & $-$0.058 & $-$0.074 \\
{\it xu$_{v}$: B}    & 0.012 & 0.287 & 0.513 & 0.200 & 1.000 & $-$0.337 & $-$0.760 & 0.510 & $-$0.084 & 0.698 & 0.498 & 0.409 & 0.507 & $-$0.256 & $-$0.095 & 0.019 & $-$0.032 \\
{\it xu$_{v}$: C}    & 0.106 & $-$0.267 & $-$0.361 & $-$0.002 & $-$0.337 & 1.000 & 0.796 & $-$0.249 & $-$0.247 & $-$0.140 & $-$0.055 & $-$0.032 & $-$0.013 & 0.092 & 0.044 & 0.026 & 0.013 \\
{\it xu$_{v}$: E}    & 0.044 & $-$0.464 & $-$0.593 & $-$0.144 & $-$0.760 & 0.796 & 1.000 & $-$0.298 & $-$0.057 & $-$0.363 & $-$0.165 & $-$0.105 & $-$0.127 & 0.133 & 0.045 & 0.024 & 0.043 \\
{\it xd$_{v}$: B}    & $-$0.049 & $-$0.054 & 0.226 & 0.048 & 0.510 & $-$0.249 & $-$0.298 & 1.000 & 0.502 & 0.437 & 0.406 & 0.344 & 0.727 & $-$0.221 & $-$0.056 & 0.014 & $-$0.056 \\
{\it xd$_{v}$: C}    & $-$0.078 & 0.196 & 0.254 & $-$0.008 & $-$0.084 & $-$0.247 & $-$0.057 & 0.502 & 1.000 & $-$0.116 & $-$0.168 & $-$0.175 & $-$0.097 & 0.107 & 0.115 & $-$0.092 & $-$0.109 \\
{\it x$\bar{U}$: C}  & $-$0.049 & $-$0.047 & 0.162 & 0.042 & 0.698 & $-$0.140 & $-$0.363 & 0.437 & $-$0.116 & 1.000 & 0.685 & 0.647 & 0.366 & $-$0.234 & $-$0.082 & $-$0.006 & $-$0.028 \\
{\it x$\bar{D}$: A}  & $-$0.098 & $-$0.140 & 0.084 & $-$0.017 & 0.498 & $-$0.055 & $-$0.165 & 0.406 & $-$0.168 & 0.685 & 1.000 & 0.961 & 0.525 & $-$0.231 & $-$0.114 & 0.049 & 0.021 \\
{\it x$\bar{D}$: B}  & $-$0.140 & $-$0.175 & 0.072 & $-$0.056 & 0.409 & $-$0.032 & $-$0.105 & 0.344 & $-$0.175 & 0.647 & 0.961 & 1.000 & 0.460 & $-$0.210 & $-$0.106 & 0.046 & 0.026 \\
{\it x$\bar{D}$: C}  & 0.018 & $-$0.369 & $-$0.100 & 0.018 & 0.507 & $-$0.013 & $-$0.127 & 0.727 & $-$0.097 & 0.366 & 0.525 & 0.460 & 1.000 & $-$0.327 & $-$0.168 & 0.133 & 0.056 \\
$a_{u}$              & 0.057 & 0.106 & $-$0.038 & 0.033 & $-$0.256 & 0.092 & 0.133 & $-$0.221 & 0.107 & $-$0.234 & $-$0.231 & $-$0.210 & $-$0.327 & 1.000 & 0.928 & $-$0.665 & $-$0.779 \\
$a_{d}$              & 0.061 & 0.093 & 0.003 & 0.057 & $-$0.095 & 0.044 & 0.045 & $-$0.056 & 0.115 & $-$0.082 & $-$0.114 & $-$0.106 & $-$0.168 & 0.928 & 1.000 & $-$0.714 & $-$0.876 \\
$v_{u}$              & $-$0.039 & $-$0.124 & $-$0.065 & $-$0.058 & 0.019 & 0.026 & 0.024 & 0.014 & $-$0.092 & $-$0.006 & 0.049 & 0.046 & 0.133 & $-$0.665 & $-$0.714 & 1.000 & 0.880 \\
$v_{d}$              & $-$0.051 & $-$0.114 & $-$0.070 & $-$0.074 & $-$0.032 & 0.013 & 0.043 & $-$0.056 & $-$0.109 & $-$0.028 & 0.021 & 0.026 & 0.056 & $-$0.779 & $-$0.876 & 0.880 & 1.000 \\

\hline
\hline
\end{tabular}
\caption{\label{tab:matrix}The correlation matrix of all parameters 
of the HH-EW-Z fit.}
\end{scriptsize}
\end{center}
\end{sidewaystable}

\newpage
\begin{table}
\renewcommand*{\arraystretch}{1.5}
\begin{center}
\begin{scriptsize}
\begin{tabular}{|l|ccc|ccc|ccc|ccc|}
\hline 
    & $a_u$ & exp & tot &  $a_d$ & exp &  tot & 
      $v_u$ & exp & tot &  $v_d$ & exp &  tot \\
\hline 
  HH-EW-Z & $+0.532$& $^{+0.081}_{-0.058}$ &$ ^{+0.107}_{-0.063}$& 
            $-0.409$& $^{+0.327}_{-0.199}$ &$ ^{+0.373}_{-0.213}$& 
            $+0.144$& $^{+0.065}_{-0.050}$ &$ ^{+0.066}_{-0.058}$& 
            $-0.503$ & $^{+0.168}_{-0.093}$ &$ ^{+0.171}_{-0.103}$ \\
\hline
ZEUS-EW-Z & $+0.50$ & $^{+0.09}_{-0.05}$ &$ ^{+0.12}_{-0.05}$& 
            $-0.56$ & $^{+0.34}_{-0.14}$ &$ ^{+0.41}_{-0.15}$& 
            $+0.14$ & $^{+0.08}_{-0.08}$ &$ ^{+0.09}_{-0.09}$& 
            $-0.41$ & $^{+0.24}_{-0.16}$ &$ ^{+0.25}_{-0.20}$ \\
\hline
\multicolumn{13}{|l|}{PDF parameters fixed to} \\
\hline
 HH-13p &   $+0.530$ & $^{+0.076}_{-0.052} $ & &
         $-0.407$ & $^{+0.313}_{-0.193} $ & &
         $+0.145$ & $^{+0.063}_{-0.050}$  & &
         $-0.500$ & $^{+0.166}_{-0.090}$  &   \\
\hline
 HERAPDF2.0& $+0.507$ & $^{+0.073}_{-0.047}$ & & 
         $-0.473$ & $^{+0.284}_{-0.166}$ & & 
         $+0.155$ & $^{+0.062}_{-0.053}$ & & 
         $-0.479$ & $^{+0.173}_{-0.110}$ & \\
\hline \hline
SM     & $+0.500$ &&& $-0.500$ &&& $+0.202$ &&& $-0.351$ && \\
\hline
\end{tabular}
\end{scriptsize}
\end{center}
\caption{\label{tab:Z}The results from HH-EW-Z
on the axial-vector and
vector couplings of the $Z$ boson to 
$u$- and $d$-type quarks.
Given are the experimental/fit (exp) and total (tot)
uncertainties.
For comparison, the results of ZEUS-EW-Z are also listed. 
In addition, results of fits with the PDFs fixed to
HH-13p and HERAPDF2.0, for which
only the couplings of the $Z$ were free parameters,
are given. 
Also listed are the SM predictions.
}
\end{table}

\newpage


\clearpage


\clearpage

\clearpage

\newpage
\newpage
\newpage


\end{document}